\documentclass[longbibliography,aps,superscriptaddress,floatfix,letterpaper,prx,reprint]{revtex4-1}

\usepackage{CJK}
\usepackage[colorlinks=true, linkcolor=cyan, citecolor=blue, urlcolor =black]{hyperref}
\usepackage{hyperref,color}
\usepackage{amsmath,mathtools}
\usepackage{bbold}
\usepackage[utf8]{inputenc} 
\usepackage{graphicx}
\usepackage{braket}
\usepackage{ragged2e}
\usepackage{wrapfig}
\usepackage{color}
\usepackage{pifont}
\usepackage{siunitx}

\usepackage[colorinlistoftodos]{todonotes}

\newcommand{\circledone}[1]{\large \ding{192} \normalsize}
\newcommand{\circledtwo}[1]{\large \ding{193} \normalsize}

\begin{document}
\begin{CJK*}{GB}{} 
\title{\sffamily Hyperbolic Lattices in Circuit Quantum Electrodynamics} 

\author{Alicia J. Koll\'{a}r}
\affiliation{Department of Electrical Engineering, Princeton University, Princeton, NJ 08540, USA}
\affiliation{Princeton center for Complex Materials, Princeton University, Princeton, NJ 08540, USA}
\author{Mattias Fitzpatrick}
\affiliation{Department of Electrical Engineering, Princeton University, Princeton, NJ 08540, USA}
\author{Andrew A.\ Houck}
\affiliation{Department of Electrical Engineering, Princeton University, Princeton, NJ 08540, USA}

\date{\today}

\begin{abstract}
After close to two decades of research and development,  superconducting circuits have emerged as a rich platform for both quantum computation and quantum simulation. Lattices of superconducting coplanar waveguide (CPW) resonators have been shown to produce artificial materials for microwave photons, where weak interactions can be introduced either via non-linear resonator materials or strong interactions via qubit-resonator coupling.
Here, we highlight the previously-overlooked property that these lattice sites are deformable and allow the realization of tight-binding lattices which are unattainable, even in conventional solid-state systems. In particular, we show that networks of CPW resonators can create a new class of materials which constitute regular lattices in an effective hyperbolic space with constant negative curvature. 
We present numerical simulations of a series of hyperbolic analogs of the kagome lattice which show unusual densities of states with a spectrally-isolated degenerate flat band. We also present a proof-of-principle experimental realization of one of these lattices. This paper represents the first step towards on-chip quantum simulation of materials science and interacting particles in curved space.
\end{abstract}

\maketitle
\end{CJK*}

\section*{\label{sec:intro}Introduction}
Euclidean space-time is the familiar geometry of non-relativistic physics. Its spatial dimensions are geometrically flat, having no inherent curvature or length scale. 
Within this paradigm, Newtonian gravity is completely described by a scalar potential.
However, this is insufficient to describe gravitational radiation or strong gravitational fields such as those found near stellar objects like black holes. These situations require a description in terms of general relativity in which gravity appears as variations in the metric of space-time.
The resulting mass-dependent curvature modifies the properties of parallel lines and geodesics, and even though particles still move along the shortest paths between points, they no longer move along the intuitive straight lines of familiar Euclidean geometry. The equations of motion must now be formulated in terms of the metric tensor, which can no longer be written in a position-independent way. The behavior of classical astrophysical objects in this mass- and position-dependent metric has been studied and simulated extensively, but a consistent formulation of both gravitation and quantum mechanics remains elusive. Therefore, a table-top simulator which naturally incorporates curvature and non-classical degrees of freedom is of considerable interest. 

Positive spatial curvature is naturally embeddable in the Euclidean laboratory frame by working with systems restricted to the surfaces of spheres. However, negative (hyperbolic) spatial curvature \cite{Cannon:1997ul} cannot be isometrically embedded in Euclidean space, so it is much more difficult to attain. To date, experimental simulations of particles in negatively curved space have been restricted to hyperbolic metamaterials in which the dielectric constant is varied to reproduce the effects of a curved metric \cite{Leonhardt:2006jl,Batz:2008bf,Smolyaninov:wa,Genov:2009fn, Chen:2010uo, Bekenstein:2017fe, Bekenstein:2015cp}. However, these experiments are purely classical and only weakly interacting. Analogs of event horizons and Hawking radiation \cite{Unruh:1980cg} have been studied both experimentally and theoretically using acoustic waves \cite{Unruh:1980cg,Weinfurtner:2011cn}, ultrashort optical pulses \cite{Philbin:2008cv}, Bose-Einstein condensates (BECs) \cite{Steinhauer:2016ih,Carusotto:2008kk}, and exciton-polariton condensates \cite{Gerace:2012ki}.
Recently methods have also been proposed for realizing the Dirac equation in curved space times using ion traps\cite{Sabin:2016fo,Pedernales:2018ep}, optical waveguides\cite{Koke:2016hq}, and optical lattices with position-dependent hopping and non-Abelian artificial gauge fields\cite{Boada:2011hd}.

Expanding upon previous work which realized Euclidean lattice models using the techniques of circuit quantum electrodynamics (cQED) and interconnected networks of superconducting microwave resonators
\cite{Houck:2012iq, Underwood:2012hx,Koch2013, Underwood:2016ju, Fitzpatrick:2017eg, Anderson:2016df}, we present a novel scheme to generate photonic materials which constitute periodic lattices in the two-dimensional hyperbolic plane \cite{Cannon:1997ul,Coxeter:1954ve}. Classical photon-photon interactions can be added by incorporating non-linear materials into the resonators, and quantum mechanical interactions by introducing qubits \cite{Houck:2012iq,Tureci2010,Raftery2014,Fitzpatrick:2017eg}.
The strongly non-classical properties of superconducting qubits and the large qubit-photon coupling rates available will allow cQED hyperbolic materials to access an entirely new regime of simulation of interacting quantum mechanics in curved space. Additionally, these systems realize much stronger curvatures than previously possible, with lattice spacings in excess of approximately $0.5$ times the curvature length.

Beyond their natural connection to general relativity, hyperbolic lattices also have significant applications in mathematics and computer science. 
Classification of these lattices and the study of their spectral properties relates to open problems in representation theory of non-commutative groups, graph theory, random walks, and automorphic forms \cite{Woess:1987wc, Sunada:1990tv, Floyd:1987up, Bartholdi:2002wb, Strichartz:1989tl, McLaughlin:1986vb, Agmon:1986tv}.

Computer scientists often study hyperbolic networks because they have several useful qualities for robust and efficient communication networks. For example, trees, which are naturally hyperbolic, are highly efficient at connecting a large number of nodes to a few central servers, and in fact the connectivity of the internet is a hyperbolic map \cite{Krioukov:2010wb, Bogu:2010cf}. 
Additionally, there exist classes of hyperbolic lattices which, unlike Euclidean lattices, cannot be split in half by the removal of a small number of nodes \cite{Lipton:1979ti}. They therefore arise frequently in the study of how to fortify a communication network against hostile tampering. This enhanced connectivity also leads to lower-overhead logical qubit encoding in surface codes \cite{Breuckmann:gs,Breuckmann:2017hy}.

Here, we will concentrate on a set of examples which are hyperbolic generalizations of the kagome lattice, and present numerical studies of their non-interacting band structures, which display highly unusual features, in particular a spectrally isolated degenerate flat band. Finally, we present experimental measurements of a device which realizes a finite section of one of these lattices, a kagome lattice made with heptagons instead of hexagons.

The remainder of this paper is organized as follows: we first introduce cQED lattices, in particular those made from 2D coplanar waveguide (CPW) resonators, and review how they can be described by tight-binding Hubbard-like effective models where the energy of ``atomic" sites is set by the resonance frequency of the microwave resonators, and the hopping rate is set by capacitive coupling between them. 
We will then highlight the previously-unnoticed fact that CPW lattice sites are naturally ``deformable'' due to their waveguide nature, allowing the realization of new types of lattices in which the inter-site spacings and the hopping rates are decoupled. We show that, in addition to opening the door to novel Euclidean lattices, this insight allows networks of CPW resonators to realize non-flat lattices, in particular hyperbolic lattices which are prohibited in other systems due to the absence of any isometric embedding even in three dimensions.
Finally, we present numerical simulations and experimental measurements of hyperbolic analogs of the kagome lattice. Further details on the effective tight-binding Hamiltonians, the numerical simulations, and lattice curvatures can be found in the Supplementary Information.

\section*{Circuit QED Lattices}\label{sec:cQEDlattices}

\begin{figure*}[!t]
	\begin{center}
		\includegraphics[width=0.9\textwidth]{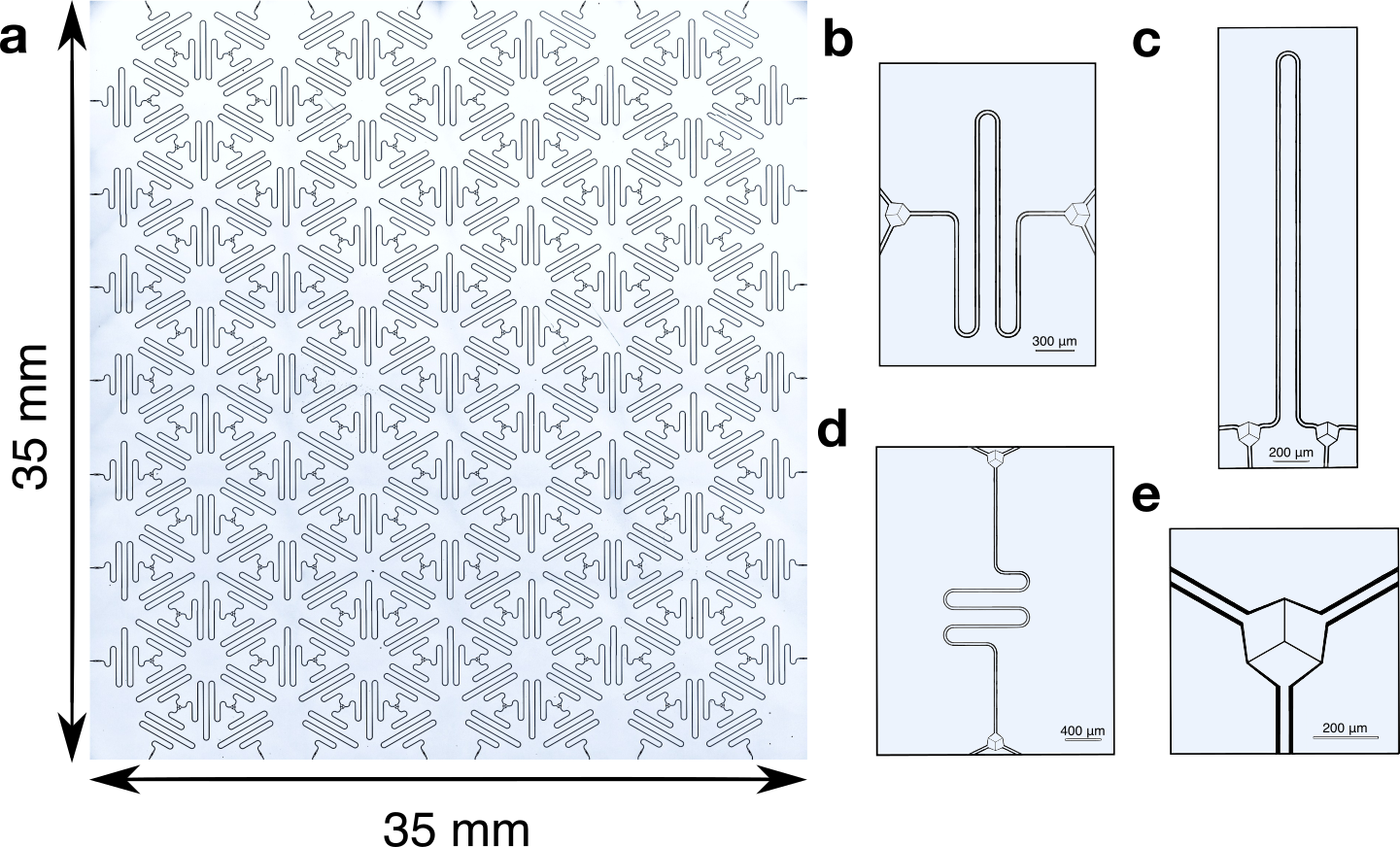}
	\end{center}
	\vspace{-0.6cm}
	\caption{\label{fig:cQEDLattices} 
    \textbf{Circuit QED lattices.} \textbf{a}, Picture of a Euclidean lattice of CPW microwave resonators, modified from \cite{Houck:2012iq}. \textbf{b}-\textbf{d}, Three mathematically identical resonators with different shapes but the same resonance frequencies and hopping rates. \textbf{e}, Close-up of a capacitive coupler like the ones used in \textbf{a}-\textbf{d} to connect three resonators together. The effective hopping rate at this junction is set by the capacitance between the three arrow-shaped centre pins.
    } 
\end{figure*}

The field of circuit quantum electrodynamics is a solid-state implementation of cavity quantum electrodynamics, in which artificial atoms, primarily superconducting qubits, are coupled to microwave resonators \cite{Blais:2004kn, Koch:2007gz, Reagor:2016hx, Anderson:2016df, Paik:2011gq}. The strong coupling regime is readily achievable, and unlike atoms, qubits are lithographically defined, so strong qubit-resonator coupling can be maintained indefinitely. This, coupled with the ease of performing high-fidelity gate operations and relatively long coherence times, has enabled superconducting qubits to emerge as a promising candidate for universal digital quantum computation. 
However, a digital, gate-based, architecture is not the only useful one. Pattern recognition and machine learning routinely use architectures based on massively interconnected neural networks.
Regular networks map naturally to lattice-based physics problem, and are therefore a convenient platform for quantum simulation of many-body physics \cite{Houck:2012iq,Koch2013}. 

Realizing this type of architecture by connecting a large number of resonators together via capacitive or inductive coupling results in artificial photonic materials. Such materials have been realized using both 2D CPW resonators \cite{Blais:2004kn,Houck:2012iq,Koch2013,Underwood:2016ju,Underwood:2016ju,Fitzpatrick:2017eg} and 3D stub resonators \cite{Paik:2011gq, Reagor:2016hx, Anderson:2016df}.

The lattices in this paper are made using CPW resonators, which can easily be etched from a single layer of superconducting film. They are a planar analog of a cylindrical coaxial cable, and consist of an electrically-isolated centre pin surrounded by two ground planes on either side. Resonators are readily defined simply by removing a section of the centre pin, and the external quality factor is set by this capacitive gap. By microfabrication standards they are quite large, several mm in length, so they are typically fabricated with meanders to compactify them, as seen in Fig. \ref{fig:cQEDLattices}b-d.

Lattices, such as the one shown in Fig. \ref{fig:cQEDLattices}a, are formed by connecting many resonators together end to end \cite{Underwood:2012hx,Underwood:2016ju,Houck:2012iq,Koch2013}. The strength of the coupling is determined by the capacitance at the junction where the centre pins converge. A close-up of a three-way junction is shown in Fig. \ref{fig:cQEDLattices}e. The hopping rate is naturally negative, and as a result, these systems exhibit band structures which are inverted from what is found in actual solid state systems, with rapidly oscillating Bloch waves at the low-energy end of the spectrum.


\begin{figure*}[!bt]
	\begin{center}
		\includegraphics[width=1.0\textwidth]{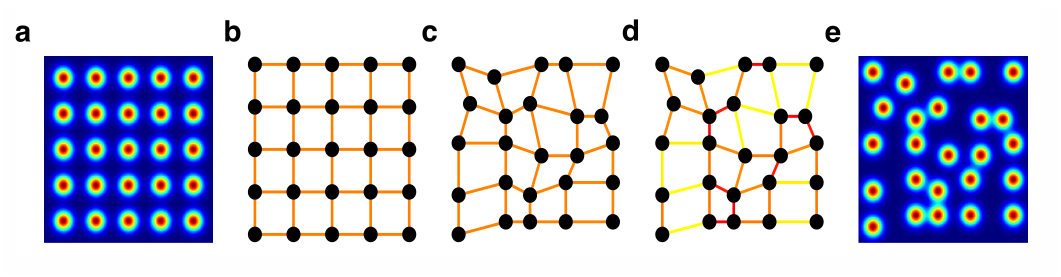}
	\end{center}
	\vspace{-0.6cm}
	\caption{\label{fig:GraphIsEverything} 
    \textbf{The graph is everything.} \textbf{a}, Colour plot of the lattice potential of a regular 2D square lattice. \textbf{b}, The corresponding regular tight-binding graph of the potential in \textbf{a}. \textbf{c}, Alternate drawing of the tight-binding graph in \textbf{a}. Despite visible displacement of the nodes, the hopping Hamiltonian is unmodified because the hopping rates, indicated by the colour of the graph edges, are identical to \textbf{b}. \textbf{d}, New tight-binding graph with the same nodes as \textbf{c} but with hopping rates dependent on the distance between the nodes. \textbf{e}, Highly disordered lattice potential which gives rise to the tight-binding graph in \textbf{d}. In systems where the effective hopping rate is determined by the distance between sites, any displacement of the lattice sites results in a disordered model with modified properties. In CPW lattices, however, the hopping rates are determined by the geometry of the coupling capacitors at each end of the resonators. In some cases, such as the curved-space lattices shown in Fig. \ref{fig:NonPlanarGraphs}\textbf{d}-\textbf{i}, the regular tight-binding graph is impossible to produce in 2D flat space, whereas a mathematically identical but distorted-looking graph like \textbf{c} can be fabricated using CPW resonators. This would not be possible in systems where hopping rates are solely determined by the distance between sites.} 
\end{figure*}

In the absence of interactions these lattices are well described by a tight-binding model, which can be written in the following form (with $\hbar=1$):

\begin{equation}\label{TBHam}
H_{TB} = \omega_{0} \sum_{i}{a_i^{\dagger}a_i} - t \sum_{<i,j>}{(a_i^{\dagger}a_j + a_j^{\dagger}a_i)},
\end{equation}
where the first sum runs over all lattice sites and accounts for the on-site energy $\omega_0$, and the second describes hopping between nearest neighbors ($\braket{i,j}$) with a characteristic rate $t<0$ \cite{Houck:2012iq,Koch2013}. The geometry of the lattice is encoded entirely in the structure of this last sum. Consider the example of a two-dimensional square lattice. A colour map of the potential is shown in Fig. \ref{fig:GraphIsEverything}a. Taking the tight-binding approximation is equivalent to replacing this continuum model with a graph-based one such as that shown in Fig. \ref{fig:GraphIsEverything}b. Each on-site wavefunction is replaced with a single complex variable that encodes its amplitude and phase. The graph has one edge for each allowed hopping transition, and they are weighted by the corresponding hopping rates. Once this assignment has been made, the tight-binding model is fully specified and, in fact, \emph{independent of how the graph is drawn}. For example, the graph shown in Fig. \ref{fig:GraphIsEverything}c is identical to that in Fig. \ref{fig:GraphIsEverything}b despite the fact that it appears disordered.

Such a deformation is challenging in an ultracold-atom or trapped-ion quantum simulator because the effective $t$ is determined by the physical distance between atoms or ions and inherently changes when they are displaced. However, the situation is quite different for CPW lattices. In this case, an individual site is a naturally one dimensional object whose connections are determined by its end points. Like a coaxial cable, a CPW resonator can be bent, and as long as the length of the resonator is fixed, its properties in the tight-binding model are unchanged. Thus, despite the differences in their physical layout, the resonators in Fig. \ref{fig:cQEDLattices}b-d are identical. Therefore, as long as the end capacitors are unchanged and the resonators can be deformed sufficiently to maintain constant total length, the tight-binding model is unchanged by moving the resonators or the locations of the couplers. \emph{It is this flexibility that makes it possible to construct curved-space lattices on flat, Euclidean substrates.}

\section*{Hyperbolic Lattices}\label{sec:HyperbolicLattices}
\begin{figure*}[!]
	\begin{center}
		\includegraphics[width=0.7\textwidth]{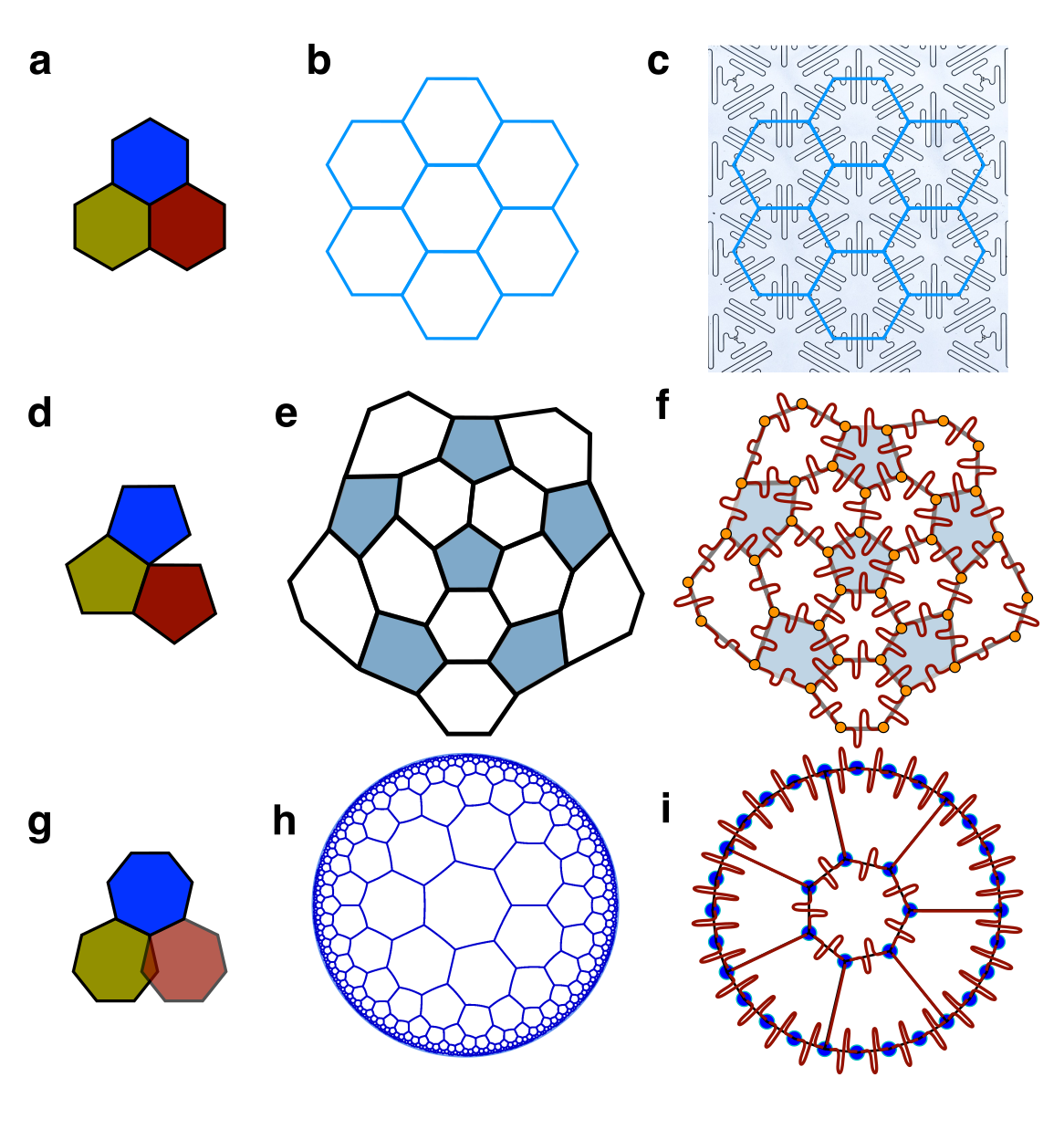}
	\end{center}
	\vspace{-0.6cm}
	\caption{\label{fig:NonPlanarGraphs} 
       \textbf{Schematic diagram of Euclidean and non-Euclidean lattices in circuit QED.}
    \textbf{a}, One vertex of a successful attempt to tile the Euclidean plane with regular hexagons.
    \textbf{b}, Resulting hexagonal lattice.
    \textbf{c}, Euclidean circuit QED lattice with the resonators laid out in regular hexagons. \emph{Because this is a valid Euclidean tiling the resonator network is highly regular, and all resonators look the same.} (Photograph modified from \cite{Houck:2012iq}.)
    \textbf{d}, One vertex of a failed attempt to tile the Euclidean plane with regular pentagons. A gap is left between the tiles, so this tiling is only valid in spherical space (positive curvature).
    \textbf{e}, Projection of a spherical soccer-ball lattice into the Euclidean plane. Some tiles must be stretched to cover the missing space.
    \textbf{f}, Schematic of a circuit QED lattice which realizes the soccer ball tiling. Resonator shapes are modified at different points in the lattice to bridge the stretched distances while preserving the hopping rates and on-site energies.
    \textbf{g}, One vertex of a failed attempt to tile the Euclidean plane with regular heptagons. The tiles overlap, so this tiling is only valid in hyperbolic space (negative curvature).
    \textbf{h}, Conformal projection of a hyperbolic heptagon lattice into the Euclidean plane.
    \textbf{i}, Schematic of a circuit QED lattice which realizes a section of the hyperbolic lattice in \textbf{h}. Resonator shapes are modified at different points in the lattice to permit tighter packing while preserving the hopping rates and on-site energies.
    } 
\end{figure*}

Typically, crystallography deals with periodic lattices consisting of a unit cell, possibly with more than one site, and a tiling of that unit cell that fills all of space with no gaps and no overlaps \cite{Ashcroft:1976ud}. Geometry therefore strongly constrains the set of all possible unit cells. However, hyperbolic and spherical polygons have larger and smaller internal angles than their Euclidean counterparts, respectively (see Fig. \ref{fig:NonPlanarGraphs}).  Therefore the set of allowed lattices is different in curved spaces. 
To see this more clearly, we will adopt a slightly non-standard approach to crystallography which generalizes to curved space more readily than the usual description in terms of unit cells and Bravais lattices. We will describe each lattice as a tiling of the plane with polygons such that each lattice site is at a vertex of the tiling, and two vertices are connected by an edge if they are nearest neighbors in the lattice. The tiling will consist of only a few distinct plaquettes, or tiles, and the geometry of the lattice can be classified in terms of the set of tiles and their tiling rule.  For example, graphene will be described as a tiling of regular hexagons such that three of them touch at every vertex, as shown in Fig. \ref{fig:NonPlanarGraphs}a.

In Euclidean space analogous tilings with three pentagons or three heptagons meeting at every vertex are forbidden. A sample vertex for each of these is shown in Fig. \ref{fig:NonPlanarGraphs}d,g. Clearly the tiles do not fit. In the pentagon case there is a gap left between the tiles, and in the heptagon case the tiles are forced to overlap. However, both of these tilings can exist in curved space \cite{Coxeter:1954ve}.

Unlike Euclidean space, spherical and hyperbolic space each have a natural length scale $R$, set by $1/\sqrt{|K|}$, where $K$ is the Gaussian curvature \cite{GRbook}. The shape of polygons in curved space depends on their size relative to $R$, and the larger they are, the more they deviate from their Euclidean counterparts \cite{Coxeter:1954ve,Cannon:1997ul}. Consequently, hyperbolic and spherical tilings can only exist for a specific size of tile, which cannot be scaled up or down without inducing gaps between the tiles or overlapping regions. Each tiling therefore has an intrinsic ratio between the curvature and the lattice spacing \cite{Coxeter:1954ve}.

Space with positive (spherical) curvature is effectively smaller than flat space. For example the circumference of a disc of radius $r$ is $2\pi R \sin(r/R) < 2 \pi r$. As a result, the gap in the tiling in  Fig. \ref{fig:NonPlanarGraphs}d can be eliminated if the curvature is sufficiently strong. This is why the famous Buckminsterfullerene \cite{Kroto:1985uw}, or ``soccer ball'' tiling of regular pentagons and hexagons can be used to cover a sphere, but not the flat plane. When such a tiling is projected flat, edges have to be stretched in order to bridge the gaps, as shown in Fig. \ref{fig:NonPlanarGraphs}e. 

Hyperbolic space with negative curvature is the opposite. It is larger than flat space, and the circumference of a disc of radius $r$ is $2\pi R \sinh(r/R) > 2 \pi r$. As a result, the overlap in the tiles in Fig. \ref{fig:NonPlanarGraphs}g can be eliminated if the curvature is strong enough, and a heptagonal analog of graphene is possible. Projecting this lattice into the flat plane requires edges to be shortened to remove overlaps, as shown in Fig. \ref{fig:NonPlanarGraphs}h. 

In order to produce an effectively curved lattice in a flat quantum simulator, this stretching or compression of the tiling must be accomplished without changing the tight-binding model. 
Schematics of realizing this transformation using CPW resonators are shown in Fig. \ref{fig:NonPlanarGraphs}f,i. In both cases the vertices of the curved tiling no longer appear in a highly regular pattern in the physical circuit, but the meanders in the resonators have been adjusted to maintain constant frequency despite the distortion, and the tight-binding model has been preserved.

The limitations of this technique are two-fold. First, the resonators must be maintained at constant frequency without overlapping on the substrate. Since the turns in a CPW cannot be made arbitrarily tight without destroying its waveguide properties, the achievable meander density cannot be made arbitrarily large. This imposes a limit on the maximum feasible distortion which typically rules out complete spherical tilings, such as Buckminsterfullerene, or radically curved hyperbolic tilings. However, finite sections of spherical tilings and of many hyperbolic lattices, such as the heptagon version of the kagome lattice, are readily achievable. 

\begin{figure}
	\begin{center}
        \includegraphics[width=0.95\columnwidth]{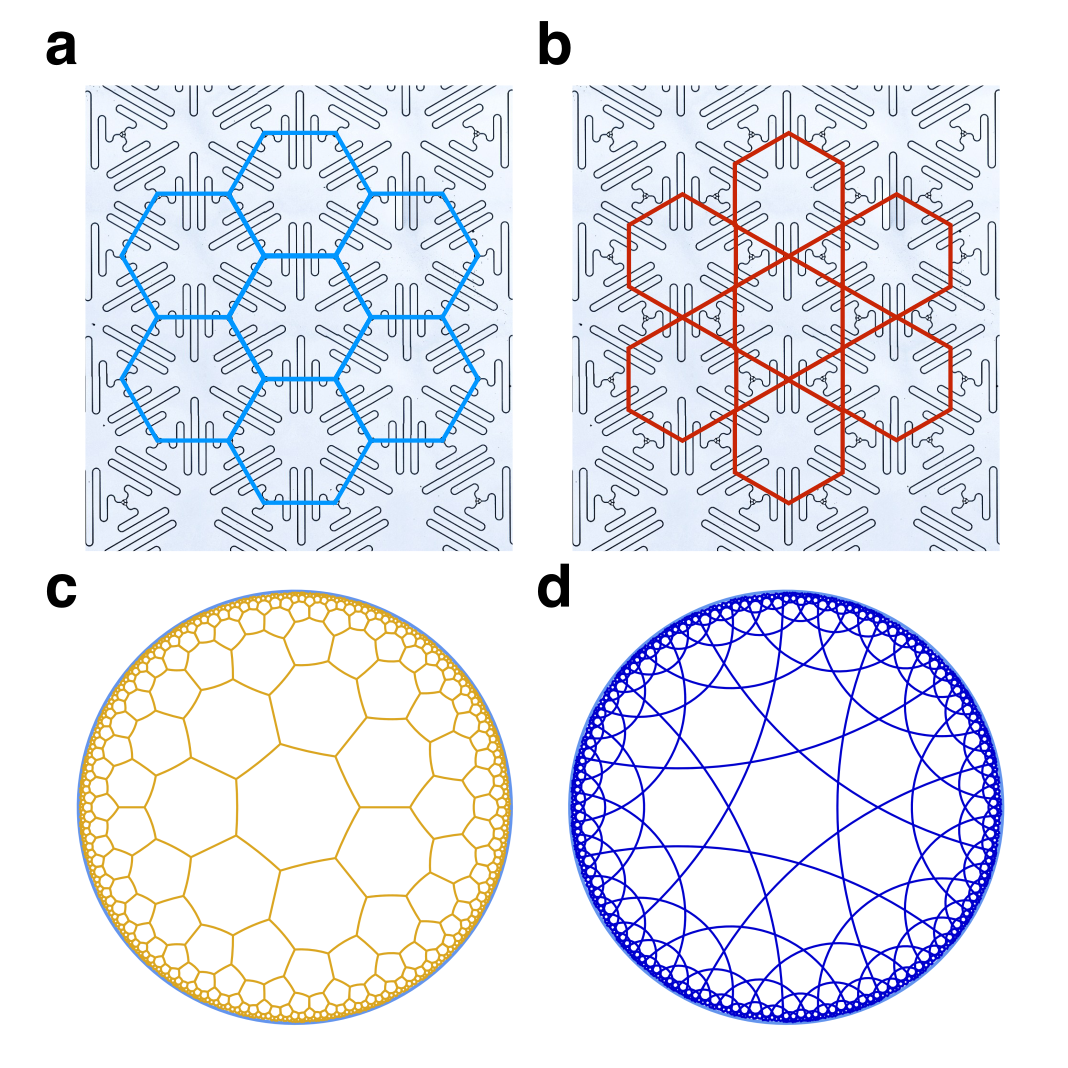}
	\end{center}
	\vspace{-0.6cm}
	\caption{\label{fig:LayoutVMedial} 
    \textbf{Layout graph vs. effective kagome graph.} \textbf{a}, Microscope image of a Euclidean lattice device with the corresponding hexagonal layout graph overlaid in blue. Resonators correspond to edges of the graph and three-way capacitors to vertices.
    \textbf{b}, The same device with the effective, kagome, tight-binding graph overlaid in red. Resonators now correspond to vertices of the graph and non-zero hopping matrix elements to the edges.
    A hyperbolic heptagonal layout graph and the resulting effective kagome-like graph are shown in \textbf{c} and \textbf{d} respectively. 
     Plots in \textbf{c} and \textbf{d} are visualized in the  Poincar\'{e} disc model \cite{Dunham:1981uk, Cannon:1997ul, Adcock:2000uw}.
     } 
\end{figure}

The second limit to the set of curved lattices achievable in circuit QED is the fact that the resonators are natively one-dimensional objects. Consequently, the most natural way to lay them down is to select a tiling and place one resonator on each edge, rather than on each vertex. The effective lattice which then appears in the tight-binding model is the medial lattice, or line graph, of the original layout, where particles live on the centre of the edges and a hopping matrix element exists between effective sites if their edges share an end point. Achievable lattices are therefore limited to those that can be written as the medial of another. Figure \ref{fig:LayoutVMedial}a-b show a Euclidean resonator lattice with an overlay of the layout lattice and effective lattice, respectively. The layout lattice is a hexagonal lattice, but the effective lattice is a kagome lattice.

The hyperbolic lattices studied in this paper are natural generalizations of the kagome lattice where the layout of the resonators is changed to a hyperbolic version of graphene, such as the heptagonal one shown in Fig. \ref{fig:LayoutVMedial}c, resulting in an effective kagome-like lattice such as that shown in Fig. \ref{fig:LayoutVMedial}d. We focus primarily on this heptagon version, which we will refer to as the heptagon-kagome lattice, because the numerical studies in the next section suggest that it belongs to a class of hyperbolic kagome-like lattices whose highly unusual band structure exhibits a spectrally isolated flat band. Additionally, the heptagon-graphene layout lattice has the weakest curvature of all possible tilings of the hyperbolic plane with a single tile, making it particularly easy to fabricate. However, even this weakest curvature is quite strong. 
The heptagon-graphene lattice has an inter-site spacing of $0.566 \, R$, and the resulting heptagon-kagome effective lattice has $0.492 \, R$ \cite{Coxeter:1954ve}.  Conversely, if we take the inter-site spacing to be that of graphene, then they would have $R = \SI{2.51}{\angstrom}$ and  $R = \SI{2.88}{\angstrom}$, respectively. Plots of other hyperbolic graphene- and kagome-like lattices and their curvatures are shown in Supplemental Fig. \ref{fig:Curvatures}.

\subsection*{Tight-Binding Simulations}\label{subsec:TB}
\begin{figure*}[!t]
	\begin{center}
		\includegraphics[width=0.80\textwidth]{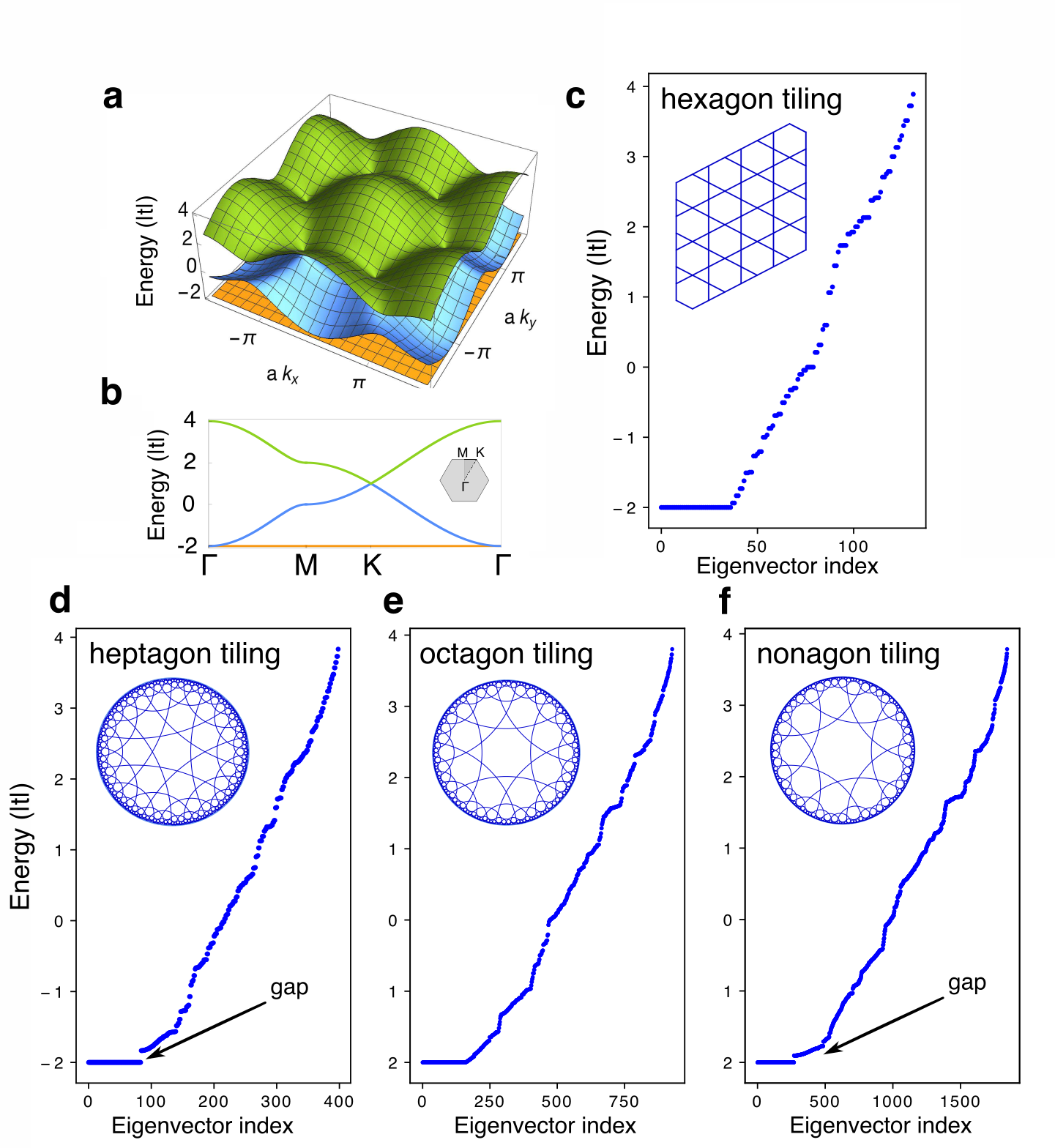}
	\end{center}
	\vspace{-0.6cm}
	\caption{\label{fig:TBNumerics} 
    \textbf{Tight-binding simulations.} \textbf{a}, Band structure of the kagome lattice for $t<0$ with periodic boundary conditions. The flat band is the lowest energy band at $E = -2|t|$. Two dispersive bands appear in the range $[-2|t|, 4|t|]$ and touch the flat band at the $\Gamma$ point at $|k| = 0$. \textbf{b}, Slice through the kagome band structure showing the band touch.
    These Bloch-theory calculations cannot be generalized to hyperbolic lattices, where the only available technique is numerical diagonalization on finite-sized lattices with hard-wall boundary conditions.
    \textbf{c}, Finite-size numerical eigenenergy spectrum for the kagome lattice. The flat band and the band touch with the dispersive bands are clearly visible, but the structure of the dispersive bands is lost.
    \textbf{d-f}, Analogous numerical eigenenergy spectra for hyperbolic versions of the kagome lattice using heptagons, octagons, and nonagons, respectively.
    The flat band is gapped for the case of odd-sided layout polygons, the heptagons and nonagons in \textbf{d} and \textbf{f}.
    } 
\end{figure*}

In Euclidean crystallography, calculation of the tight-binding band structure of a lattice is straightforward once the unit cell, Bravais lattice, and hopping rates have been determined. Translation groups in Euclidean space are commutative, so representation theory guarantees that a Bloch-wave ansatz will yield eigenstates and eigenenergies as a function of momentum \cite{Ashcroft:1976ud}.

In hyperbolic space, however, the discrete translation groups are non-commutative. As a result, there is no natural analog of a Bravais lattice. In fact, simply writing down the locations of all the lattice sites and the directions of all the bonds is already a non-trivial mathematical problem. To date, no hyperbolic equivalent of Bloch theory exists, and there is no known general procedure for calculating band structures in either the nearly-free-electron, or tight-binding limits. 
Specialized methods are known for the cases of trees \cite{Chen:1974bn}, but fail if there are any closed loops, except in the special case of Cayley graphs of the free products of cyclic groups \cite{McLaughlin:1986vb}.
The only universal method is numerical diagonalization of the hopping Hamiltonian. This is a brute force method which yields a list of eigenvectors and eigenvalues, but no classification of eigenstates by a momentum quantum number. It, therefore, cannot be used to directly obtain the dispersion relations. Instead, it is more useful as a measure of the energy spectrum and the density of states (DOS).

We produce finite-size samples using a cylindrical construction in which we start with a single layout polygon and successively add shells of nearest-neighbor polygons. We compute the effective lattice of this truncated layout, neglecting all hopping matrix elements that would connect to resonators outside the simulation size, and obtain a hopping matrix in a localized, delta-function, basis. Numerical diagonalization of this matrix yields eigenenergies and eigenstates for each lattice type and simulation size. A sample energy spectrum for a three-shell piece of the Euclidean kagome lattice is shown in Fig. \ref{fig:TBNumerics}c. It has significantly reduced information compared to the full band structure calculation plotted in Fig. \ref{fig:TBNumerics}a-b, but the presence of the flat band, and accompanying delta-function spike in the DOS, at $-2|t|$ is clear. Spectra for the hyperbolic heptagon-, octagon-, and nonagon-kagome lattices are shown in Fig. \ref{fig:TBNumerics}d-f. All four spectra share several similar features: a flat band at $-2|t|$ and the remaining, presumably dispersive, bands filling the range from $-2|t|$ to $4|t|$. For the two tilings formed from odd-sided polygons a spectral gap is visible between the flat band and the remaining eigenstates. We have verified that this gap is independent of system size and decreases with the number of sides of the layout polygon. The other gaps visible in the spectrum appear to be finite-size artifacts that close with increasing system size.
(See Supplemental Fig. \ref{fig:SSize} for details.)

Flat bands, like those seen in Fig. \ref{fig:TBNumerics}, are quite rare in band structures \cite{Bergman:2008es,Leykam:2018vd}. However, because the kinetic energy in these bands  effectively vanishes, they are ripe for strongly correlated many-body physics and non-perturbative interactions, such as fractional Quantum Hall states arising from discrete Landau levels. However, a pure flat-band description is only valid if the interaction strength is smaller than the gap to the nearest dispersive band. This is very much the case with Landau levels where magnetic energies far outweigh tunneling energies and tight-binding band structure, but otherwise gapped flat bands are highly unusual. Among Euclidean lattices there are only a handful of known examples where real-space topology of the flat band allows a gap to exist \cite{Bergman:2008es,Leykam:2018vd}.
However, there are entire classes of hyperbolic systems that display gapped flat bands, including the $(2n+1)$-gon-kagome lattices, and certain generalizations of trees.

In all cases, the flat band arises because of an infinite multiplicity of localized eigenstates. In the kagome lattice the smallest such state consists of a single occupied hexagonal loop with alternating sign of the wavefunction on each site and vanishing occupation elsewhere. This state can easily be generalized to the octagon-kagome lattice, but not to the heptagon or nonagon versions where the odd number of sides introduces geometric frustration. In these cases, the flat band still consists of localized states in the form of an alternating closed loop, but it must now extend over two tiles. See Supplemental Information and Supplemental Fig. \ref{fig:ManyEigenstates}g for details. 
Without a hyperbolic generalization of Bloch theory, the real-space topology argument which proves that the flat band of the kagome lattice is not gapped \cite{Bergman:2008es} cannot be readily extended to hyperbolic kagome lattices.
As a result, while the flat-band eigenstates have been identified in all cases, the origin of the spectral gap isolating these states from the rest of the spectrum for $(2n+1)$-gons remains unknown.

\subsection*{Experiment}\label{subsec:DeviceMeasurements}
\begin{figure*}[!t]
	\begin{center}
		\includegraphics[width=0.85\textwidth]{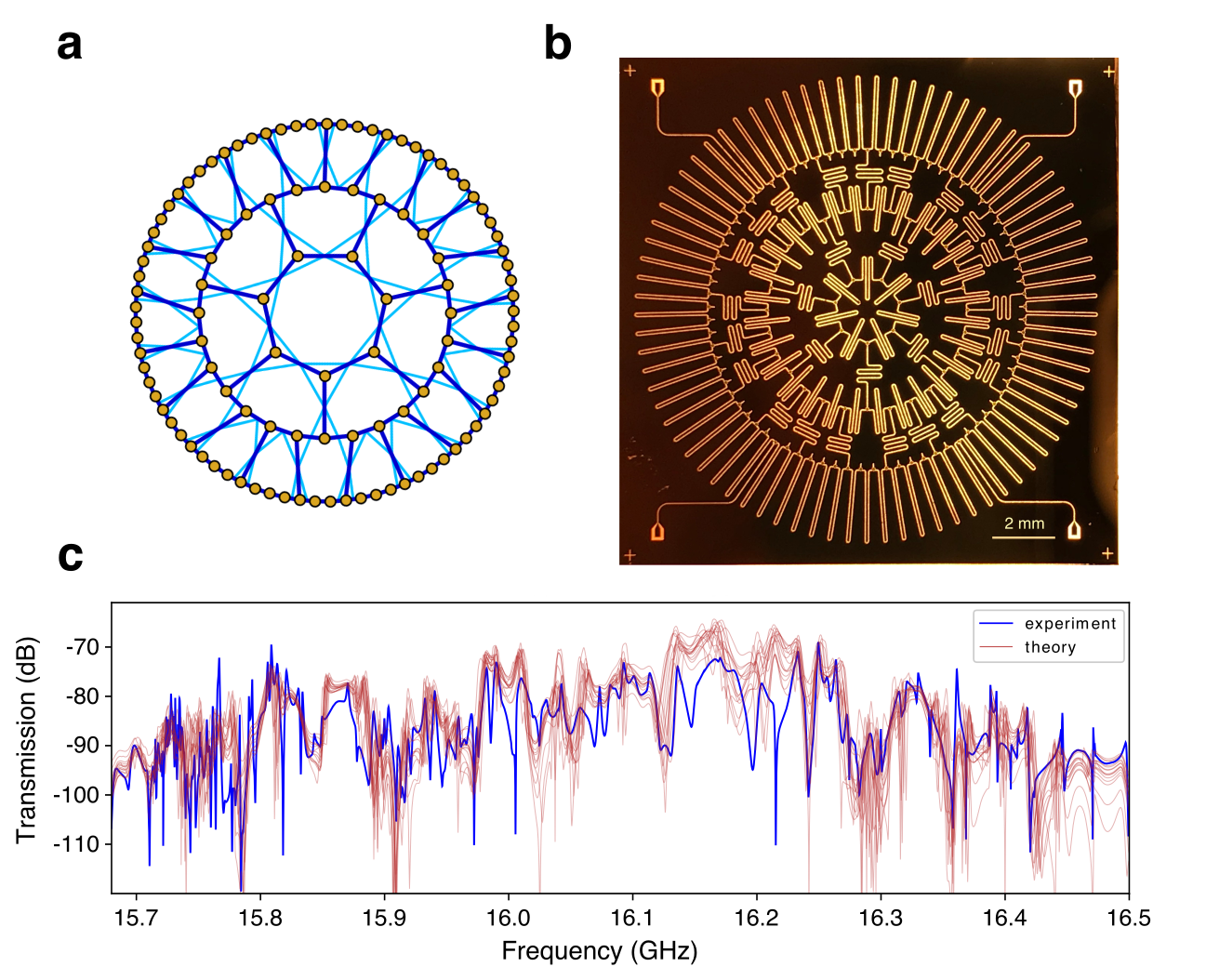}
	\end{center}
	\vspace{-0.6cm}
	\caption{\label{fig:HKDevice} 
    \textbf{The heptagon-kagome device.} \textbf{a}, Resonator layout (dark blue) and effective lattice (light blue) for a circuit that realizes two shells of the heptagon-kagome lattice. Orange circles indicate three-way capacitive couplers. \textbf{b}, Photograph of a physical device which realizes the layout and effective graphs in \textbf{a}. The device was fabricated in a 200 nm niobium film on a sapphire substrate and consists of 140 CPW resonators with fundamental resonance frequencies of 8 GHz, second harmonic frequencies of 16 GHz, and a hopping rate of $-136.2 \ \rm{MHz}$ at the second harmonic. Four additional CPW lines have been included at each corner of the device to couple microwaves into and out of the device for transmission measurements. 
    Short stubs protruding inward from the outermost three-way couplers in the device are high frequency $\lambda/4$ resonators which maintain a consistent loading of the sites in the outer ring--ensuring uniform on-site energies. 
    \textbf{c}, Experimental transmission ($S_{21}$) for the device in \textbf{b} is shown in dark blue. The red curves show theoretical transmission for an ensemble of theoretical models including small systematic offsets in the on-site energies and realistic disorder levels, demonstrating reasonable agreement between theory and experiment. 
    } 
\end{figure*}

We have constructed a device to realize a finite section of the heptagon-kagome lattice. It consists of one central heptagon and two shells of neighboring tiles. A schematic of the layout is shown in Fig. \ref{fig:HKDevice}a, where each resonator has been approximated by a single line, and the lengths have not been held fixed. The actual device is fabricated using photolithography to etch CPW resonators into a $200$ nm film of niobium on a 500 $\mu$m sapphire substrate. A photograph of the device is shown in Fig. \ref{fig:HKDevice}b. The resonators are ~$7.5$ mm long with a fundamental resonance frequency of $8$ GHz and a second harmonic of $16$ GHz. The second harmonic of this device realizes the heptagon-kagome lattice with a hopping rate of $-136.2$ MHz. In order to minimize parasitic systematic frequency differences between resonator geometries, each resonator type was fabricated individually and the corresponding resonant frequencies were measured. 
Commercial microwave simulation packages were unable to achieve the required level of absolute or relative accuracy, so
the resonator lengths were then fine-tuned empirically to remove the residual offsets at the level of $30$ MHz. For the device shown in Fig. \ref{fig:HKDevice}b, the average difference between the fundamental frequencies of resonators with different shapes is approximately 0.13\%(10 MHz), limited by intrinsic reproducibility within a fabrication run\cite{Underwood:2012hx} and wire-bonding or parasitic capacitances sensitive to variations between fabrication runs. 
Each individual shape has a fabrication induced reproducibility of 0.036\%(2.9MHz), consistent with previous work\cite{Underwood:2012hx}. In addition to the lattice itself, the circuit also contains four measurement ports, visible in each corner, used to interrogate the lattice.

In order to simulate the transmission properties of the device, we use the tight-binding calculations described above, including systematic errors in the on-site energies and small additional levels of random disorder in both the on-site energies and hopping rates.  After numerical diagonalization, we obtain the frequency and wavefunction of each eigenmode of the lattice. Assuming a typical HWHM of $1.4 \ \rm{MHz}$, we generate a Lorentzian resonance profile for each mode, centred about its eigenenergy. We then compute the level of mode matching between the eigenmodes and the input and output ports using the numerical wavefunctions. To calculate the transmitted electric field versus probe frequency, we sum the transmission through all $140$ eigenmodes, weighted by their Lorentzian line shapes and spatial mode matching, 
and add an empirical background due to leakage through the device packaging. (See Methods for details.)
Theoretical transmission curves for 15 different disorder realizations are shown in Fig. \ref{fig:HKDevice}c,
along with a plot of the experimental transmitted power near the second harmonic frequency of the device.
(The fundamental modes of the device obey a different tight-binding model due to the asymmetry of the mode function within each resonator \cite{Koch2013}. See the Supplementary Information for details.)
These theoretical curves reproduce most of the qualitative features of the data including: the onset of peaks, the location and Fano-like lineshapes of the highest frequency peaks, and the markedly larger linewidth of the modes near $16.2$ GHz which have the largest overlap with the coupling ports.
This device therefore demonstrates that hyperbolic lattices can be produced on chip using CPW resonators, and paves the way to the study of interactions in hyperbolic space, and to simulation of new models with non-constant curvature.

\section*{Conclusion}\label{sec:conclusion}
In conclusion, we have shown that circuit QED lattices of two-dimensional CPW resonators can be used to produce artificial photonic materials which exist in an effective curved space. In particular, we conducted numerical tight-binding simulations of a class of hyperbolic analogs of the kagome lattice, and demonstrated that they display a flat band similar to that of their Euclidean counterpart. However, for the case of odd-sided polygons this band is isolated from the rest of the spectrum. We also constructed a proof-of-principle experimental device which realizes a finite section of non-interacting heptagon-kagome lattice. Mathematical investigation into the origin of the gap is still ongoing.
In addition to the curved-space lattices discussed here, this technique can be used to create a large variety of other graphs and lattices such as two-dimensional sheets with ripple distortions and Cayley trees.

While our present work is purely non-interacting, interactions can be included via incorporating kinetic inductance materials such as NbTiN into the resonators to obtain a classical $\chi^3$ non-linearity \cite{Annunziata:2010dg,Rotzinger:2014tl}, or via the addition of qubits to each resonator. These methods may also be applied to computer science or mathematics problems such as the study of non-linear operators on trees or Cayley graphs of free products of cyclic groups, rather than the lattices discussed here.
Alternatively, it is possible to appropriate the techniques of hyperbolic metamaterials in which the equations of motion are tailored to mimic the existence of a nontrivial metric by deliberate modulation of the dielectric constant. A discrete version of the same effect\cite{Koke:2016hq,Boada:2011hd} can be realized here by tailoring the hopping magnitude, providing a simple route to models of more moderate curvature and specific metrics such as the Schwarzschild solution. The promise of strong interactions in these lattices leads to an exciting frontier which may provide answers to questions at the interface of quantum mechanics, gravity, and condensed matter physics, as lattices with these properties cannot be fabricated from actual materials.

\section*{methods}

The device was fabricated using photolithography and reactive ion etching to pattern CPW resonators in a $200$ nm film of niobium on a 500 $\mu$m sapphire substrate. The resonators were $7.5$ mm long with a target fundamental resonance frequency of $8$ GHz and a second harmonic of $16$ GHz, and respective hopping rates of $-70$ MHz and $-140$ MHz. 
In addition to the resonator lattice itself, the device also includes the four external coupling ports visible in each corner of Fig. \ref{fig:HKDevice}b.
Initially, the length of all resonators was designed constant throughout the device by varying the number and extent of the meanders, in order to compensate for the required changes in the distance between the end points of each resonator.
However, it was found that initially there were small residual differences in frequency (up to $0.37\%$ or $30$ MHz) for resonators of different shapes. 

In order to compensate for this each unique resonator shape was fabricated on different chips and measured individually. Then, incremental changes were made to the total length to eliminate the parasitic frequency errors. The u-shaped resonators on the outer ring were the most numerous type, so all other resonators were adjusted to match their measured frequency of $7.99$ GHz. 
The test resonators used to compensate for the resonator frequency offsets also provided an independent measure of intrinsic disorder levels. Within a single fabrication run resonators of the same geometry had variations in their frequencies of $0.036\%$ ($2.9$ MHz), which is consistent with previous results \cite{Underwood:2012hx}. Compensation for the systematic offsets between shapes was limited by the intrinsic reproducibility and variations between fabrication runs. For most of the resonator shapes, the systematic disorder in the on-site energy was less than  $0.13\%$ ($10$ MHz). However, the middle and outer rings of resonators proved more sensitive to fabrication variations, presumably due parasitic capacitances in the long parallel straight sections of the waveguides. Systematic offsets in these resonators could be eliminated to the $0.19\%$ ($15$ MHz) level.

The final, disorder-compensated, device was mounted in a microwave printed circuit board using indium seals and aluminum wire bonds. 
The device was anchored to the base plate of a dilution refrigerator and the input to the lattice was provided using a high-frequency microwave generator. 
The fundamental modes of the device constitute a heptagon-kagome lattice with mixed sign of the hopping matrix elements (See Supplemental Information details). We therefore measure the symmetric second-harmonic modes near $16$ GHz using a cryogenic HEMT amplified and heterodyne down conversion.

The connectivity and on-site energies of the device were verified using theoretical reconstructions of expected transmission spectra.
Eigenmodes and eigenenergies were computed numerically assuming an intrinsic linewidth and a set of on-site energies. Each mode was assumed to have an external quality factor proportional to its overlap with the four coupling ports and a Lorentzian lineshape.
Summing these simulated lineshapes produced a theoeretical reconstruction of the expected electric field transmitted through the lattice.

In the frequency range of $15$-$17$ GHz the main limitation to measurement is leakage transmission through the PCB housing around the device, which gives rise to a frequency-depended coherent background in all transmission measurements with variations on the scale of $50$-$100$ MHz. Unfortunately, this background is extremely sensitive to the exact microwave boundary conditions around the device and cannot be measured independently of the device. However, it is significantly more broadband than the device transmission peaks. We therefore use the experimental data to gain an empirical estimate of this background signal. Using a Gaussian convolution filter, we remove the components of the amplitude signal which vary rapidly with frequency. The resulting filtered signal constitutes an experimental measure of the leakage power.
The phase of the device transmission swings by many multiples of $2\pi$, making it impossible to unwrap and perform a similar reconstruction of the phase of the leakage signal. We therefore estimate the leakage electric field by taking the square root of the leakage power, and neglect the slow phase variations that are present in the actual device.

Adding this empirical background to the theoretical electric field obtained from numerical diagonalization yields a simulated transmission curve. We then input the average resonator frequency, the hopping rate, and the systematic offsets in the on-site energies, allowing a different average value for each azimuthal ring and each of the two sets of radial resonators. In order to further match the device, we introduce disorder in the device parameters at the level given by the individual resonator tests and produce ensembles of such theoretical curves.  Figure \ref{fig:HKDevice}c shows the experimental transmission curve in blue and a best-fit  ensemble of fifteen theoretical simulations in red. 
The theoretically estimated on-site energies are within estimated error. The two outer rings are slight outliers with systematic offsets of $-7$ and $+15$ MHz at the fundamental modes, whereas all the other shapes lie within $\pm 5$ MHz. The theoretically estimated hopping rate of $-136.2$ MHz is consistent with software simulations.

\vskip 0.2in\vskip 0.2in
\begin{acknowledgments}
We thank Peter Sarnak, J\'{a}nos Koll\'{a}r, Rivka Bekenstein, Charles Fefferman, and Siddharth Parameswaran for fruitful discussions.
This work was supported by the NSF, the Princeton Center for Complex Materials DMR-1420541, and by the MURI W911NF-15-1-0397.
\end{acknowledgments}


%

\newpage

\section*{Supplementary Information}
\subsection*{CPW Resonators and Circuit QED Lattices}
The CPW microwave resonators used in this work are formed by electrically isolating a portion of the centre pin. Networks are formed by butting the ends of the centre pins up against each other, as shown in Fig. \ref{fig:cQEDLattices}e. Larger coordination numbers than three are possible, but the circuit layout tends to produce significant next-nearest-neighbor hopping which greatly complicates the description of the network, and will not be discussed here.

When two resonators are coupled together, their modes hybridize and produce bonding and anti-bonding configurations. Which of these is lower in frequency is determined by the energetics of the electric field distribution given the type of coupling element. When the resonators are laid out in a periodic lattice the circuit naturally maps to a tight-binding model of a 2D crystalline solid. The type of lattice, square, kagome, etc., is determined by how the resonators are connected together, and the frequency of the resonator maps to the atomic binding energy of the conduction band. The capacitive coupling between resonators can be expressed as a nearest-neighbor hopping rate, $t$, whose magnitude is set by the size and shape of the gap between the centre pins. The sign of the matrix element must be chosen to yield the correct physical behavior given the sign of the on-site wavefunctions.

The eigenmodes of the CPW resonators are standing waves with an antinode of the voltage at each end of the cavity. Quantum computing applications typically use the fundamental, half-wave, mode. However, when studying network or lattice physics, the inherent asymmetry of the half-wave mode complicates the sign of the hopping rates. In many cases this minus sign can be removed by a local gauge transformation which redefines positive and negative on alternating sites \cite{Koch2013}. However, for many of the hyperbolic lattices considered here, this gauge transformation is not possible, and it is impossible to consistently define all the hopping rates with the same sign. This complicates the mathematical description and, in some cases, induces additional geometric frustration. The second-harmonic, or full-wave, mode is symmetric and does not have this sign problem. In the main text of this paper, we therefore restrict to full-wave modes, and all results, unless explicitly stated, will refer to this case.

Regardless of whether the on-site wavefunction is even or odd, it is energetically favorable for the voltages on either side of the coupling capacitors connecting the resonators to be of opposite sign. Circuit QED lattices therefore exhibit band structures which are inverted from what is found in typical solid state-systems, with rapidly oscillating Bloch waves at the low-energy end of the spectrum. For full-wave modes, this physics can be captured by a standard tight-binding model with negative $t$. For half-wave modes, a global definition of the sign of $t$ may no longer be possible, and care must be taken \cite{Ashcroft:1976ud,Koch2013}. We proceed by defining an orientation for each resonator and arbitrarily designate one end of each resonator ``positive'' and one end ``negative''. For the tight-binding model to be physical the sign of $t$ must be consistent with the designated orientation and always favor opposite voltages on either side of the coupling capacitors. Therefore, $t<0$ when it describes hopping between two ends of the same sign and $t>0$ when describing hopping between two ends of different sign.

As hinted above, the choice of the orientation of each half-wave resonator is arbitrary and merely a bookkeeping convention. However, some choices are, of course, more convenient and illustrative than others. For some layout lattices, e.g. square or hexagonal, the orientation of each resonator can be chosen consistently such that all hopping matrix elements are negative, even for half-wave modes \cite{Koch2013}. In these cases, the asymmetry of the on-site wavefunction can be gauged away by this judicious choice of orientations, and the half-wave tight-binding model is isomorphic to the version with even on-site wavefunctions, in much the same way that the magnon spectrum of an unfrustrated antiferromagnet is identical to that of the corresponding ferromagnet. However, if the layout lattice contains plaquettes with an odd number of sides, then an orientation which fixes the sign of the hopping matrix elements cannot generally be found. Compared to their full-wave counterparts, these tight-binding models exhibit additional geometric frustration due to the phase windings forced by the half-wave mode function.

This effect can most easily be seen by considering the maximally excited state of a kagome-like layout lattice, which has eigenvalue $4|t|$. The defining features of this state are: uniform probability of occupation of all the sites and  equal voltages on either side of all coupling capacitors. For full-wave modes, both of these conditions are easily satisfied by uniform occupation of all cavities with a single phase. For half-wave modes, this is not the case. Consider a single layout plaquette of the lattice and define the orientations such that going from the positive to the negative end of a resonator corresponds to moving clockwise around the plaquette. In order maintain equal voltages on both sides of all coupling capacitors using half-wave modes, the sign of the on-site wavefunction must alternate on neighboring sites, which cannot be done consistently on a plaquette with an odd number of sides.

\begin{figure*}[!t]
	\begin{center}
		\includegraphics[width=0.9\textwidth]{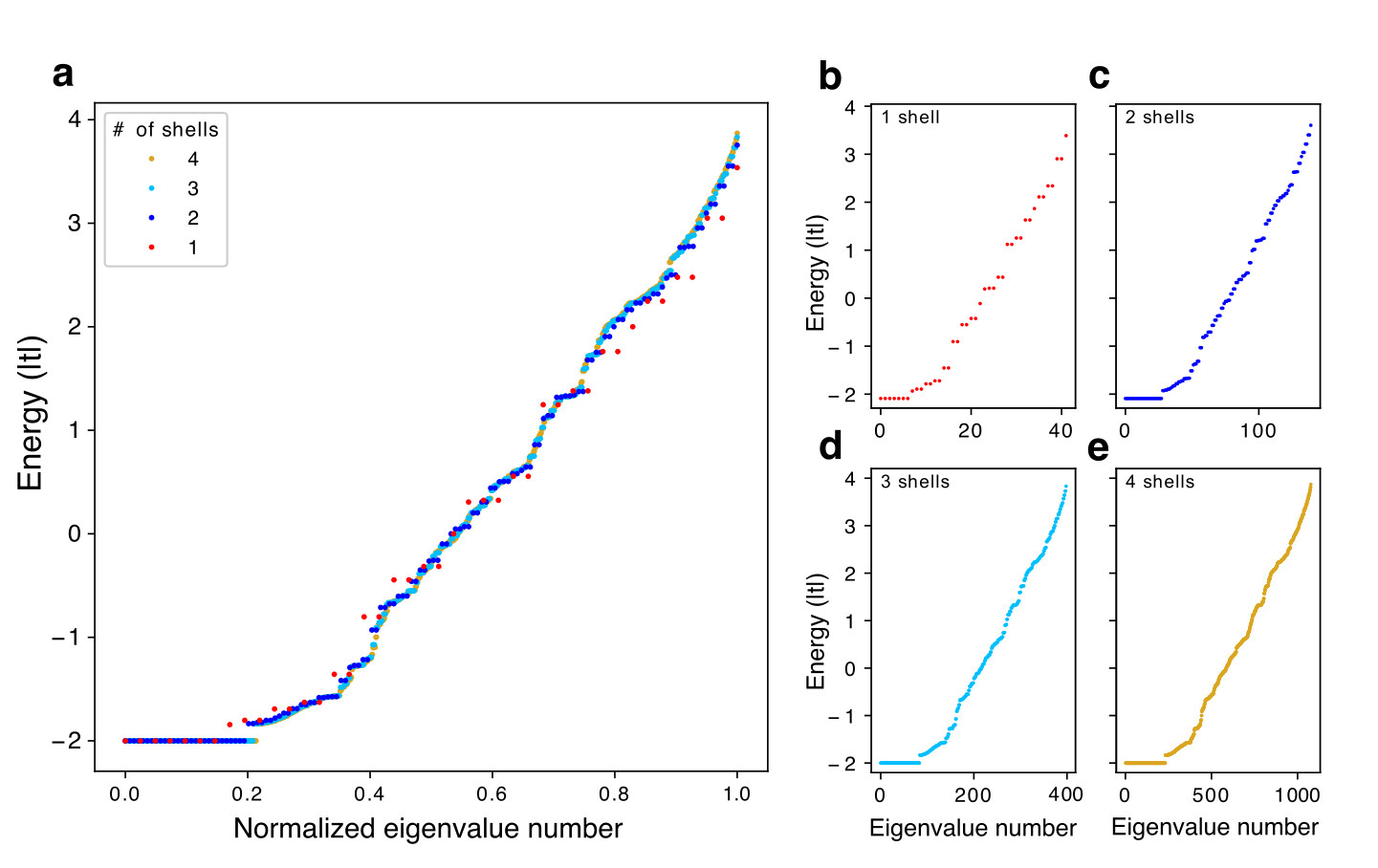}
	\end{center}
	\vspace{-0.6cm}
	\caption{\label{fig:SSize} 
    \textbf{System-size effects.} \textbf{a}, Numerical eigenenergy spectra for the heptagon-kagome lattice versus system size. The smallest layout tiling consists of a central heptagon and a shell of seven immediate neighbors (1 shell). Each successive tiling includes another shell of immediate neighbors of the previous one. The spectra are plotted for one (red), two (blue), three (cyan), and four (gold) shells of neighbors. The results for each individual system size are shown in \textbf{b}-\textbf{e}.
    The density of states converges rapidly with number of shells despite relatively small system sizes and effective hard-wall confinement.
    The theoretical plots elsewhere in this paper are from three-shell simulations. The experimental device consists of  two shells.
    } 
\end{figure*}

Hyperbolic kagome lattices made from even-sided polygons are free of this effect and the asymmetry of the half-wave mode function can always be gauged away. Kagome-like lattices made from odd-sided polygons, on the other hand, are frustrated. Remarkably, the flat-band is immune to these frustration effects because it is spanned by states which consist of loops with an even number of edges, such as the one shown in Fig. \ref{fig:ManyEigenstates}g. The dispersive bands above $E= -2|t|$, however, are affected, and the spectral gap which isolates the flat band in the full-wave case disappears from the half-wave model.

\subsection*{Tight-Binding Models: System Size Effects}

Numerical diagonalization studies of finite-size tight-binding models of the heptagon-kagome lattice were conducted for a series of system sizes, starting with a smallest simulation consisting of a central layout heptagon and one shell of nearest-neighbor plaquettes. We compute the effective lattice of this finite layout, neglecting all links to sites outside the current system size. Larger systems are created iteratively by adding successive shells of nearest neighbor layout polygons to the edge of the smaller simulations and recomputing the effective lattice.

Numerical eigenenergy spectra for a series of system sizes are shown in Fig. \ref{fig:SSize}. The flat band is clearly visible even for the smallest simulation with only one shell of neighboring polygons. The spectral gap between the flat band the rest of spectrum is already present, but not readily distinguishable from the finite-size-induced gaps at higher energy. Within the size of simulation currently possible, the higher-lying gaps all close progressively with increasing system size, and it remains unclear whether any of them also remain open at infinite size. Since the spectral gap between the flat band and the rest of the energy spectrum is readily visible for systems with two or more layers of neighbor polygons, our physical device, shown in Fig. \ref{fig:HKDevice}, was constructed at this depth.

\subsection*{Numerical Eigenstates}
\begin{figure*}[!t]
	\begin{center}
		\includegraphics[width=0.9\textwidth]{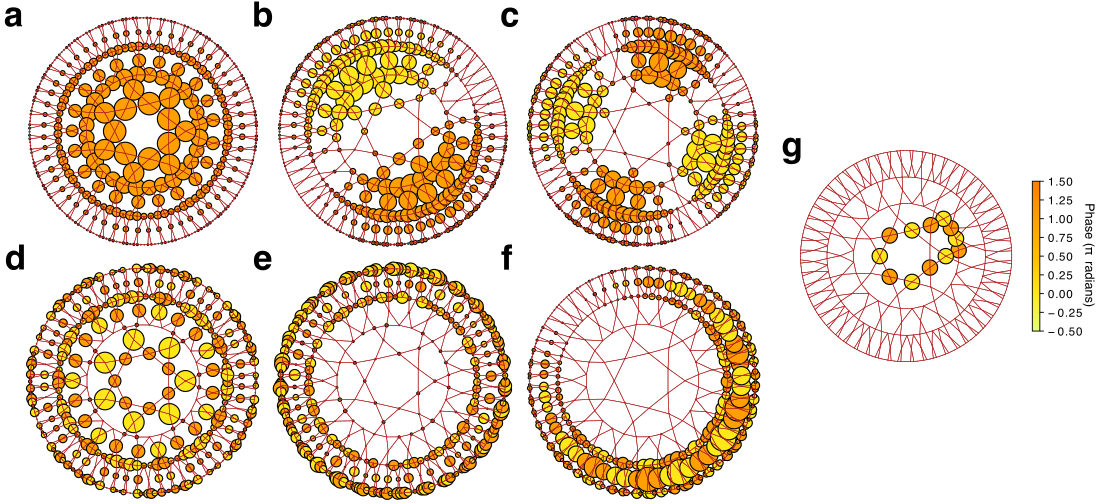}
	\end{center}
	\vspace{-0.6cm}
	\caption{\label{fig:ManyEigenstates} 
    \textbf{Colour plots of selected numerical eigenstates of three shells of the heptagon-kagome lattice.} States are ordered from highest to lowest energy, and are  plotted by placing a circle on each lattice site of the effective lattice. The size of the circle indicates the amplitude of the state on that site, and the colour its phase. \textbf{a}, The maximally excited state. This state is uniform in phase, but its amplitude varies radially due to the effective confinement from the missing links at the boundary of the simulation. \textbf{b}-\textbf{c}, Examples of the two next-highest states. They bear a striking resemblance to Laguerre-Gaussian or particle-in-a-cylindrical box modes found in flat Euclidean space. \textbf{d}-\textbf{f}, Selected intermediate excited states. Notice that the state in \textbf{f} shows both amplitude and phase modulation in the azimuthal direction, with independent periods. 
    \textbf{g}, The localized eigenstate of compact support that forms the flat band.
    } 
\end{figure*}

Colour plots of selected numerical eigenstates from a three-shell simulation of the heptagon-kagome lattice are shown in Fig. \ref{fig:ManyEigenstates}. Due to the negative hopping amplitude, the uniform-phase configuration typically associated with the ground state of a lattice is, in fact, the highest excited state, shown in Fig. \ref{fig:ManyEigenstates}a. Additionally, it exhibits an radial amplitude modulation due to confinement from the missing links beyond the edge of the sample, which constitute an effective hard-wall potential. The next eigenstates down, Fig. \ref{fig:ManyEigenstates}b,c, bear a striking resemblance to the Laguerre-type 0,1 and 0,2 modes of a Euclidean harmonic oscillator or a particle in a cylindrical box. The qualitative similarity persists for many modes throughout the spectrum, such as the one shown in Fig. \ref{fig:ManyEigenstates}d which resembles a Laguerre 4,7 mode. 

Conversely, the eigenstates also show several features which are strikingly different from the Euclidean case. Because of the hyperbolic nature of the lattice, a macroscopic fraction of the lattice sites are in the outermost ring, and the spectrum therefore contains a relatively large number of states like the one in Fig. \ref{fig:ManyEigenstates}e which reside primarily at the edge of the system. Another intriguing feature is the existence of states like the one in Fig. \ref{fig:ManyEigenstates}f, which display azimuthal amplitude and phase modulation with different periods.

A flat-band eigenstate is shown in Fig. \ref{fig:ManyEigenstates}g. The analogous state in the kagome lattice encircles a single hexagonal plaquette, but this type of behavior is not possible in a hyperbolic kagome-like lattice formed with odd-sided polygons because the sign flips cannot be consistently maintained. Therefore, the flat-band state in the heptagon kagome lattice (or any $2n+1$-gon kagome lattice) consists of a single loop across two plaquettes in which the phase flips by $\pi$ between every neighboring pair of sites. 
It is a localized eigenstate which is protected from hopping by destructive interference in the triangular plaquettes which border the loop. Since this state has compact support, translations of it are orthogonal and form a degenerate manifold of states whose multiplicity is proportional to the system size.

\subsection*{Lattice Curvatures}
\begin{figure*}[!t]
	\begin{center}
		\includegraphics[width=0.76\textwidth]{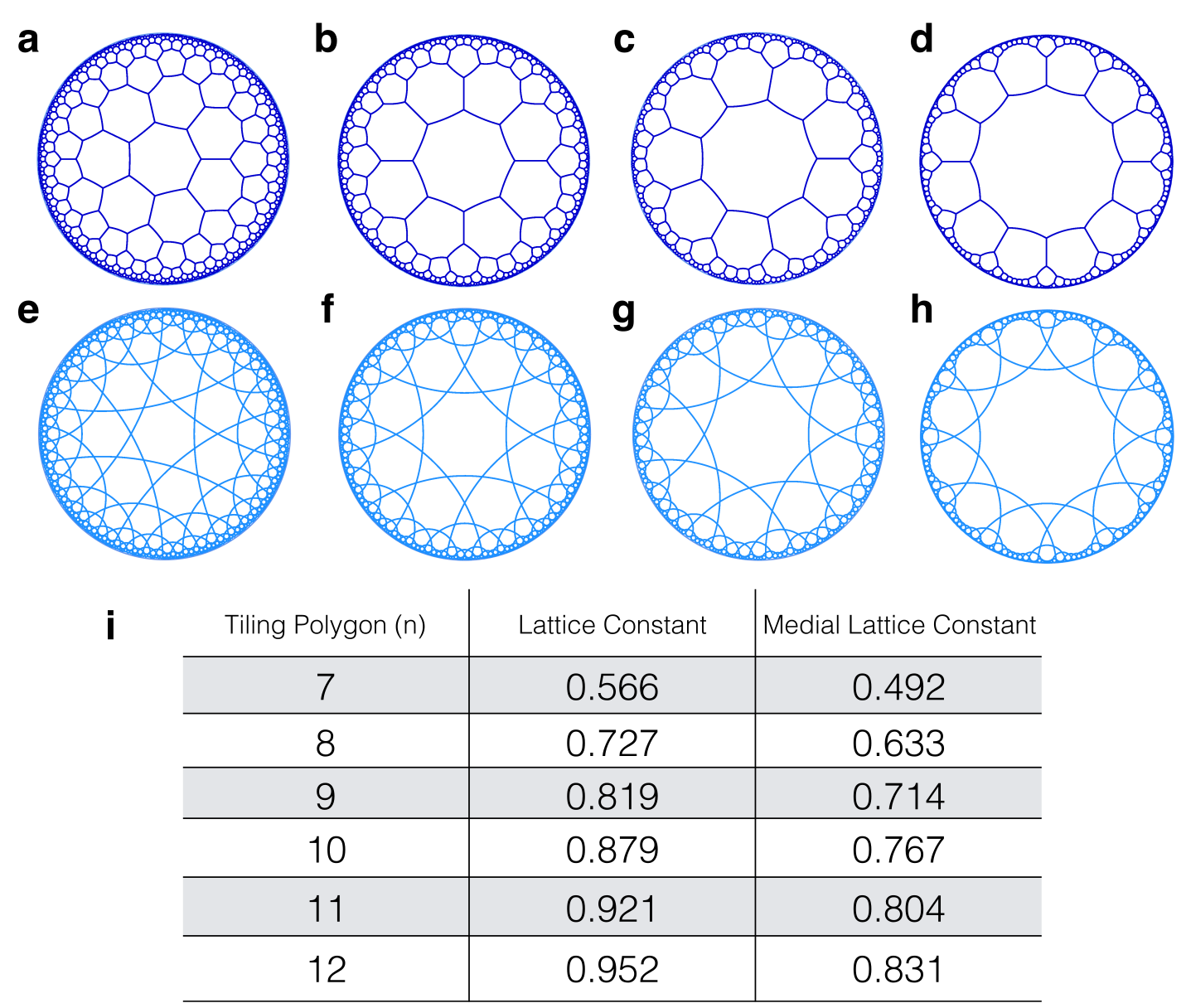}
	\end{center}
	\vspace{-0.6cm}
	\caption{\label{fig:Curvatures} 
    \textbf{ Poincar\'{e}-disc-model conformal projections of hyperbolic lattices.} \textbf{a}-\textbf{d}, Graphene-like lattices formed from heptagons, octagons, nonagons, and dodecagons, respectively. \textbf{e}-\textbf{h}, The corresponding kagome-like effective lattices which arise when \textbf{a}-\textbf{d} are used as the layout lattice. \textbf{i}, A table of inter-site spacings for graphene-like hyperbolic layout lattices and their medial lattices, the kagome-like effective lattices. All distances are given in terms of the curvature length $R = 1/\sqrt{|K|}$. 
    As the number of sides of the layout polygon increases, the intrinsic curvature of the tiling also grows, and the polygons become markedly larger.
    } 
\end{figure*}


Compared to spherical or hyperbolic space, familiar Euclidean space is unique in that there is no intrinsic length scale. In particular, polygons have the same internal angles, regardless of size. Therefore, any Euclidean tiling can be scaled up or down in size and remain unchanged. This is not true in spherical or hyperbolic space where there is a natural length scale $R$ which is set by the Gaussian curvature $K = \pm 1/R^2$ \cite{GRbook}. In both cases, the shape of a polygon depends on its size relative to this length scale \cite{Coxeter:1954ve,Cannon:1997ul}. For example, consider a triangle on the surface of a sphere which has one vertex at the north pole and the other two on the equator separated by $\pi/2$ longitudinally. The total internal angle of this triangle is $3\pi/2$ (each angle is $\pi/2$), but if it is reduced in size until it is much smaller than the radius of curvature of the sphere, it tends to a Euclidean triangle with a total internal angle of $\pi$. Tiles in spherical space therefore get ``fatter'' as they increase in size and can cover gaps left between Euclidean tiles of the same shape. Hyperbolic polygons, however, have smaller internal angles at their vertices than their Euclidean counterparts. Therefore, hyperbolic tilings are precisely those for which the tiles would overlap if drawn according to Euclidean geometry. (See Fig. \ref{fig:NonPlanarGraphs}.) As hyperbolic polygons become larger and larger, their internal angles decrease progressively, and they become ``pointier''.  

Because of this size-dependent geometry, whether or not a set of tiles can form a valid lattice in curved space depends on their size. The polygons must be precisely big enough so that the tiles fit exactly. Increasing the size of the polygons any further will cause gaps or overlaps to appear between the tiles, depending on the sign of the curvature. Therefore, for fixed curvature, each non-Euclidean lattice can only exist for a specific tile size. Conversely each lattice can be though of as having an intrinsic curvature given by the required ratio of inter-site spacing and $R$.

In order to determine these curvatures for the hyperbolic tilings considered in this paper we make use of the Poincar\'{e} disc model conformal mapping of the two-dimensional hyperbolic plane with curvature $-1$ onto the Euclidean unit disc \cite{Cannon:1997ul}. For concreteness, we consider only hyperbolic tilings which are generalizations of graphene to polygons with a larger number of sides. These are precisely those that produce kagome-like effective models. The layout tiling will exist when a hyperoblic $n$-gon can be produced which is bounded by geodesics and whose hyperbolic internal angle is $2\pi/3$, allowing copies of it to be tiled in a graphene-like way with three tiles meeting at every vertex. Since the Poincar\'{e} disc model is a conformal model which preserves angles, it suffices to determine the polygon size which results in an internal angle of $2\pi/3$ when drawn in the unit disc. The geodesics of the Poincar\'{e} disc model are circles which intersect the unit disc at an angle of $\pi/2$ and the metric is known analytically, so the radius and edge length of the tile required to form the layout lattice can be readily determined. The corresponding inter-site spacing for the effective lattice is determined by computing the distance between the midpoints of the edges of the layout polygon. 

For the heptagon-graphene layout lattice the inter-site spacing is determined to be $0.566 R$ \cite{Coxeter:1954ve}, and the curvature of the resulting heptagon-kagome effective lattice is slightly weaker with an inter-site spacing of $0.492 R$. Lattice spacing and Poincar\'{e} disc model plots of a series of graphene- and kagome-like lattices are shown in Fig. \ref{fig:Curvatures}. The ratio of the lattice spacing to $R$ increases with increasing number of sides of the layout polygon because the corresponding attempt at a Euclidean tiling has more and more overlap that needs to be removed by increasing the size of the tiles. This is clearly visible in the plots as the polygon at the origin occupies a larger and larger fraction of the unit disc. In fact, the heptagon-graphene tiling has the smallest ratio of lattice spacing to curvature of any tiling of the hyperbolic plane by a single polygon, regardless of vertex coordination number. Correspondingly, the heptagon-kagome has the weakest curvature of all the resulting effective lattices, making this pair natural entry points to the study of hyperbolic lattices.

\end{document}